\documentclass{optica-article}

\journal{opticajournal} 

\articletype{Research Article}

\usepackage{lineno}

\begin{document}

\title{Atmospheric Turbulence-Resilient Long-Range Fourier Ptychography}

\author{Junhao Zhang,\authormark{1,2,3,4} Weilong Wei,\authormark{1,3,4} Kaiyuan Yang,\authormark{1,2,3,4} Qiang Zhou,\authormark{2,5,6} Haotong Ma,\authormark{1,3,4} Ge Ren,\authormark{1,3,4} and Zongliang Xie\authormark{1,3,4,*}}

\address{\authormark{1}The Institute of Optics and Electronics, Chinese Academy of Sciences, Chengdu 610209, China\\
\authormark{2}University of Electronic Science and Technology of China (UESTC), Chengdu 610054, China\\
\authormark{3}Key Laboratory of Optical Engineering, Chinese Academy of Sciences, Chengdu, 610209, China\\
\authormark{4}University of Chinese Academy of Sciences, Beijing 100049, China\\
\authormark{5}Institute of Fundamental and Frontier Sciences, University of Electronic Science and Technology of China, Chengdu 610054, China\\
\authormark{6}Research Center for Quantum Internet, Tianfu Jiangxi Laboratory, Chengdu 641419, China}

\email{\authormark{*}xiezl@ioe.ac.cn}

\begin{abstract*} 
While Fourier ptychography (FP) offers super-resolution for macroscopic imaging, its real-world application is severely hampered by atmospheric turbulence, a challenge largely unaddressed in existing macroscopic FP research operating under idealized conditions. This work establishes, to our knowledge, the first comprehensive computational framework specifically designed for turbulence mitigation in long-range FP, termed Turbulence-Mitigated FP (TMFP). Rather than correcting pupil errors, an image degradation model is developed alongside a reconstruction pipeline inspired by speckle interferometry. By taking multiple short-exposure randomly-distorted measurements and exploiting their statistical properties, the diffraction-limited sub-aperture images can be recovered for further FP reconstruction. Numerical simulations and experimental validations under optical turbulence demonstrate the method's robustness, resolution enhancement, and practicality in adverse conditions, paving the way for the reliable deployment of high-resolution macroscopic FP in real-world scenarios.

\end{abstract*}

\section{Introduction}
Spatial resolution of practical optical systems faces two fundamental limitations: aperture size and atmospheric turbulence. Aperture size governs the diffraction limit, as described by the Rayleigh criterion, restricting the minimum resolvable feature size with the angular resolution given by $1.22\lambda/D$, where $\lambda$ is the center wavelength and $D$ denotes the size of imaging aperture. Meanwhile, when the aperture size surpass the atmospheric coherence length, the system's angular resolution transitions from being aperture-limited to turbulence dominated regime. For example, atmospheric turbulence limits angular resolution to about 0.5-1 arcsec; consequently, a 4-m telescope has approximately the same resolving capabilities as a small 10-cm diameter tube\cite{liu_research_2025}. These intertwined physical limitations hinder high-resolution imaging in scenarios ranging from ground-based astronomy to airborne surveillance.

Synthetic aperture (SA) imaging attempts to circumvent the diffraction barrier inherent to single-aperture systems\cite{miller_optical_2007} by synthesizing a large aperture with multiple sub-apertures, the resolution is then determined by the synthetic aperture size, which can exceed that of a single aperture by orders of magnitude. However, traditional interferometric SA techniques, such as those used in radio astronomy\cite{moreira_tutorial_2013,ryle_synthesis_1960}, rely on precise phase coherence across distributed receivers but face challenges in optical regimes as the phase information will be lost\cite{miller_optical_2007,xie_15-m_2023,yang_model-driven_2024}. An optical SA system must ensure the confocality (spatial overlap of focal volumes) and cophase alignment (phase synchronization) among sub-apertures, which necessitates ultrahigh-precision phase detection and dynamic posture control, hindering its widespread engineering applications. 

Fourier ptychography (FP) has emerged as a promising computational SA technique\cite{zheng_wide-field_2013,zheng_concept_2021}. It merges the principle of phase recovery and SA imaging, achieving sub-diffraction imaging without strict phase matching requirements. FP reconstructs high-resolution complex fields by iteratively stitching low-resolution images captured under varied illumination angles or different aperture location. Since its first demonstration\cite{zheng_wide-field_2013}, FP has gained substantial research interest, finding significant applications in biomedical and pathological studies\cite{ou_quantitative_2013,dong_high-resolution_2014,nguyen_deep_2018,zheng_concept_2021,song_synthetic_2022,zhou_fourier_2023,wu_lens-free_2024,zhao_deep-ultraviolet_2024,zhang_toward_2024}. In parallel, efforts have been directed towards extending FP principles to macroscopic scales for long-range imaging\cite{dong_aperture-scanning_2014,holloway_toward_2016,holloway_savi_2017,li_farfield_2023,article,tian_optical_2023,wang_learning-based_2023,li_snapshot_2024,xiang_coherent_2021,xiang_phase_2022}. In 2014, Dong \emph{et al.}\cite{dong_aperture-scanning_2014} demonstrating that by replacing angle-varied illumination with aperture or camera scanning, FP can be extended to far-field imaging as macroscopic FP. Subsequent advancements by Holloway \emph{et al.}\cite{holloway_toward_2016} introduced a synthetic aperture visible imaging (SAVI) framework based on macroscopic FP, followed by a long-range reflective FP implementation demonstrating a sixfold resolution enhancement\cite{holloway_savi_2017}. The recent work have focused on improving macroscopic FP's performance, including expanding the imaging field of view\cite{li_farfield_2023,article}, reducing speckle noise\cite{tian_optical_2023}, improving temporal resolution\cite{wang_learning-based_2023,li_snapshot_2024}, and correcting aberrations\cite{xiang_coherent_2021,xiang_phase_2022}.

However, these advancements predominantly assume ideal, turbulence-free conditions, creating a significant gap between laboratory demonstrations and practical field deployment. Atmospheric turbulence disrupts imaging through spatiotemporal fluctuations in refractive index $(\Delta n(\boldsymbol{r},t))$, governed by Kolmogorov's $\kappa^{-11/3}$ power law\cite{noll_zernike_1976,harding_fast_1999}. Over long propagation paths, accumulated phase distortions $(\phi_{turb}=\frac{2\pi}{\lambda}\int\Delta nds)$ fragment the optical wavefront into speckle patterns and reduce the effective coherent aperture area to $\sim(D/r_0)^2$ isoplanatic patches governed by Fried's parameter $r_0$\cite{fried_optical_1966}. For FP, which relies critically on coherent overlap between adjacent sub-aperture spectra for phase retrieval, this manifests as two fatal disruptions: (1) dynamic phase errors disrupt the phase consistency required for sub-aperture stitching, and (2) each captured intensity becomes a turbulence-corrupted realization, violating the FP forward model. While prior work \cite{paxman_phase-diversity_1994} models turbulence as a time-varying pupil error, existing FP aberration correction techniques (e.g., pupil function recovery\cite{xiang_coherent_2021} and phase diversity\cite{xiang_phase_2022}) fail here as they are designed for static errors and lack the temporal resolution required for dynamic turbulence compensation.

In the field of astronomy, computational image recovery have demonstrated the power of recovering diffraction-limited resolution through dynamic disturbance, known as speckle interferometry\cite{labeyrie_attainment_1970}. This technique exploits inherent statistical properties of distorted image ensembles to separately recover both Fourier amplitude and phase of the hidden object. Despite its efficiency in astronomic imaging\cite{denker_near_2001}, its potential in long-range FP remains unexplored. 

In this work we report, to our knowledge, the first comprehensive computational framework specifically designed for turbulence mitigation in long-range FP, termed Turbulence-Mitigated FP (TMFP), that integrates the principle of speckle interferometry into FP acquisition and reconstruction process. For image acquisition, the key experimental modification of TMFP is capturing multiple short-exposure images at each aperture scanning position, which allows us to utilize the statistical properties of random distortions. For image reconstruction, TMFP works in a dual-stage pipeline: (1) mitigating turbulence-induced distortions by fusing multiple short-exposure measurements using speckle interferometry, reconstructing diffraction-limited images with suppressed artifacts; (2) synthesizing the high-resolution image from the turbulence-mitigated data via Fourier ptychographic reconstruction. This approach maintains the relative simplicity of the FP hardware setup (requiring only extended camera exposure time per position), and leverages computational efficiency. Numerical and experimental validations through optical turbulence demonstrate TMFP’s effectiveness, robustness, and applicability to real-world scenarios. In the following sections, we first introduce the theoretical foundation and algorithmic implementation of the proposed method. We then validate our method with numerical and experimental results. Finally, we summarize the results and discuss future directions based on the proposed method.

\section{Method}

\subsection{Forward formation model}

\begin{figure*}[ht]
	\centering
	\includegraphics[width=\linewidth]{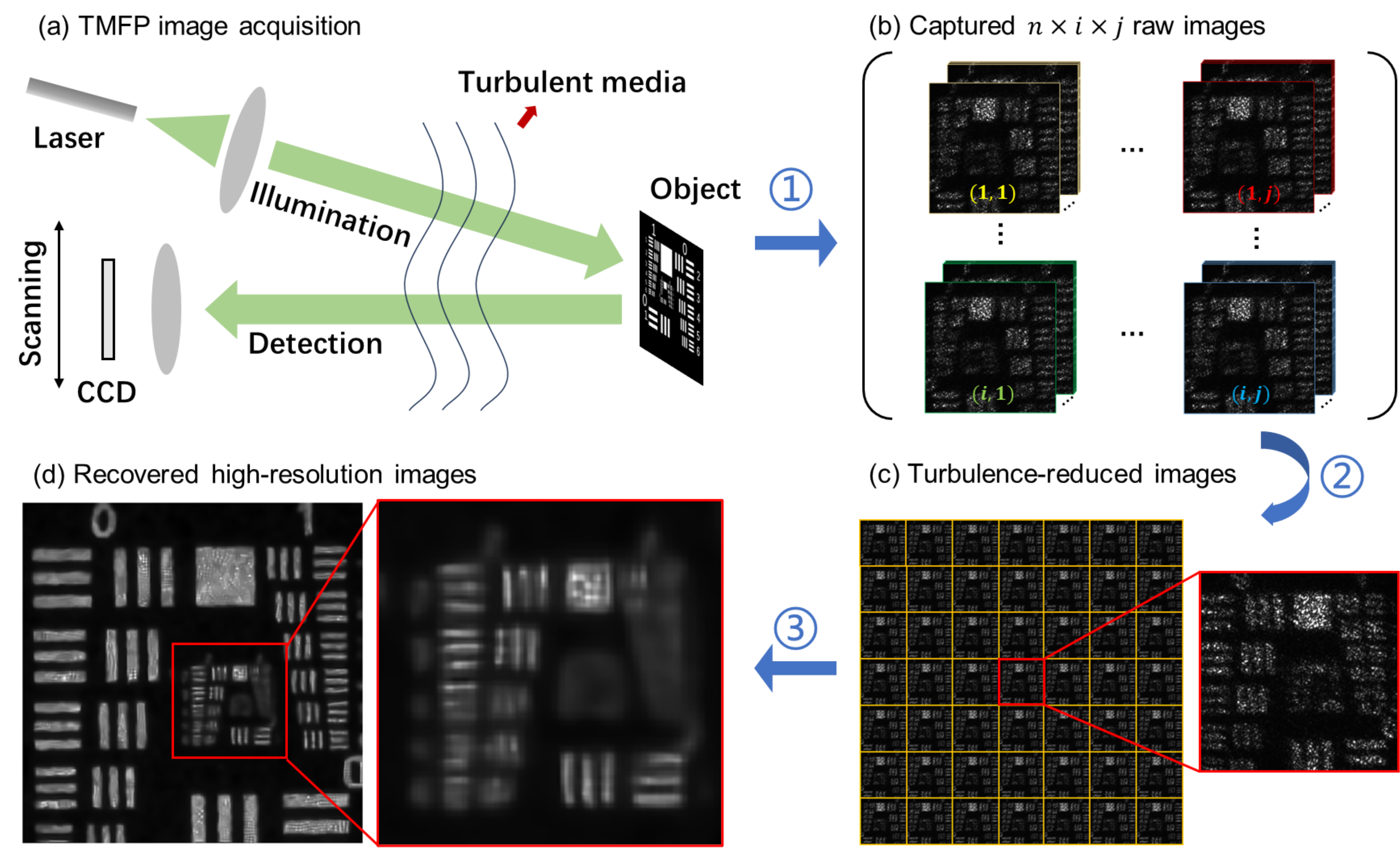}
	\caption{Principle of turbulence-mitigated Fourier Ptychography (TMFP). (a) A distant object is illuminated with a coherent source, the reflected signal is captured by a camera through turbulent media. The camera is scanning over the desired synthetic aperture. In each scanning location, $n$th short-exposure random-distorted raw images are captured (Step 1). (b) By translating the camera to $i\times j$ positions, a total of $n\times i\times j$ images are captured. Coherent averaging is performed for raw images at each scanning location (Step 2), then we obtain (c) turbulence-corrected data which is composed of $i\times j$ recovered diffraction-limited images. After computational phase retrieval and speckle denoising (Step 3), we finally obtain (d) the recovered high-resolution images.}
	\label{fig:fig1}
\end{figure*}

The forward formation model is depicted in Figure \ref{fig:fig1}a. The TMFP system employs a coherent laser source and a projective lens to illuminate the target, with reflected light captured by a translating camera to acquire low-resolution (LR) images across multiple scanning positions. While sharing the core architecture of conventional macro-FP systems, TMFP introduces a critical innovation: at each scanning position, multiple short-exposure images are captured to sample turbulence-distorted realizations, as shown in Figure \ref{fig:fig1}b. The forward model can be summarized as:
\begin{equation}
	I_{n}(\boldsymbol{x},\boldsymbol{c})\propto |\mathcal{F}\{\Omega(\boldsymbol{u}-\boldsymbol{u_c})\cdot P_{n}(\boldsymbol{u})\}|^{2},
	\label{eq:eq1}
\end{equation}
where $\boldsymbol{x}$ and $\boldsymbol{u}$ denote the spatial and spatial-frequency coordinates, respectively, with $\boldsymbol{u_c}$ representing the relative frequency shift induced by the $\boldsymbol{c}$-th aperture position. The complex wavefield emanating from the object is described by $\Omega (\boldsymbol{u})$, and the pupil function is modeled as $P_{n}(\boldsymbol{u})$, which defines the coherent transfer function in the Fourier domain and acounts for the $n$-th unique turbulence distortion. The Fourier transform operator is denoted by $\mathcal{F}\{\cdot\}$. The $n$-th captured intensity image, corresponding to the $c$-th aperture position, is expressed as $I_{n}(\boldsymbol{x},\boldsymbol{c})$, where the subscript $n $ indexes sequential acquisitions under turbulence-distorted conditions at the same frequency spectrum location. In turbulent environments, these images are severely corrupted by turbulence-induced phase distortions, rendering conventional FP reconstruction ineffective. While phase distortion can be viewed as pupil error, it is inherently time-varying. However, conventional FP can only corrects static pupil errors and is therefore unlikely to be suitable for real-time correction. Phase diversity\cite{xiang_phase_2022} represents a promising approach, but it introduces additional system complexity.

\subsection{Principle of turbulence mitigation}

The principle of TMFP is based on the idea that the diffraction-limited sharp image (FP’s raw data) can be recovered from multiple random-distorted short-exposure images, by extracting their hidden information. As stated before, the recovery the diffraction-limit image in speckle interferometry is divided into two steps: (1) recovering the Fourier amplitude and (2) recovering the Fourier phase. However, while this method is based on incoherent imaging model, FP is essentially based on a coherent imaging model. To bridge this gap, we model distorted images as convolution products between the original sharp image $I(\boldsymbol{x})$ and unknown random blur kernels $h_{n}(\boldsymbol{x})$, expressed as:
\begin{equation}
	I_{n}(\boldsymbol{x}) = I(\boldsymbol{x})\otimes h_n(\boldsymbol{x}),
	\label{eq:eq2}
\end{equation}
and
\begin{equation}
	\tilde{I}_{n}(\boldsymbol{u}) = \tilde{I}(\boldsymbol{u})\cdot\tilde{H}_n(\boldsymbol{u}),
	\label{eq:eq3}
\end{equation}
where $\otimes$ denotes the convolution operator, $I_{n}(\boldsymbol{x})$ denotes the $n$th distorted image, Equation (\ref{eq:eq3}) is the Fourier transform of Equation (\ref{eq:eq2}). The established image degradation model is now formally consistent with the incoherent imaging model. We aims to recover the original sharp image $I(\boldsymbol{x})$ from the distorted image $I_{n}(\boldsymbol{x})$, which is equivalent to recover its Fourier spectrum $\tilde{I}(\boldsymbol{u})$. To recover the amplitude of $\tilde{I}(\boldsymbol{u})$, Labeyrie\cite{labeyrie_attainment_1970} first demonstrated that short-exposure images retain high-spatial-frequency information up to the diffraction limit, despite turbulence-induced phase corruption. By ensemble-averaging the power spectra $<|\tilde{I}_{n}(\boldsymbol{u})|^{2}>$ of multiple distorted images, the original image's Fourier amplitude $|\tilde{I}(\boldsymbol{u})|^{2}$ can be retrieved as:
\begin{equation}
	<|\tilde{I}_{n}(\boldsymbol{u})|^{2}>=|\tilde{I}(\boldsymbol{u})|^{2}\cdot<|\tilde{H}_n(\boldsymbol{u})|^{2}>.
	\label{eq:eq4}
\end{equation}

Here, $<|\tilde{H}_n(\boldsymbol{u})|^{2}>$ is the energy spectrum of the turbulent point spread function (PSF), which describes how the spectral components of the image are transmitted by the atmosphere and the imaging system. At every moment this function is unknown but its time-averaged value can be determined as non-zeros up to the diffraction limit under stationary turbulence conditions.

\begin{figure}[ht]
	\centering
	\includegraphics[width=10cm]{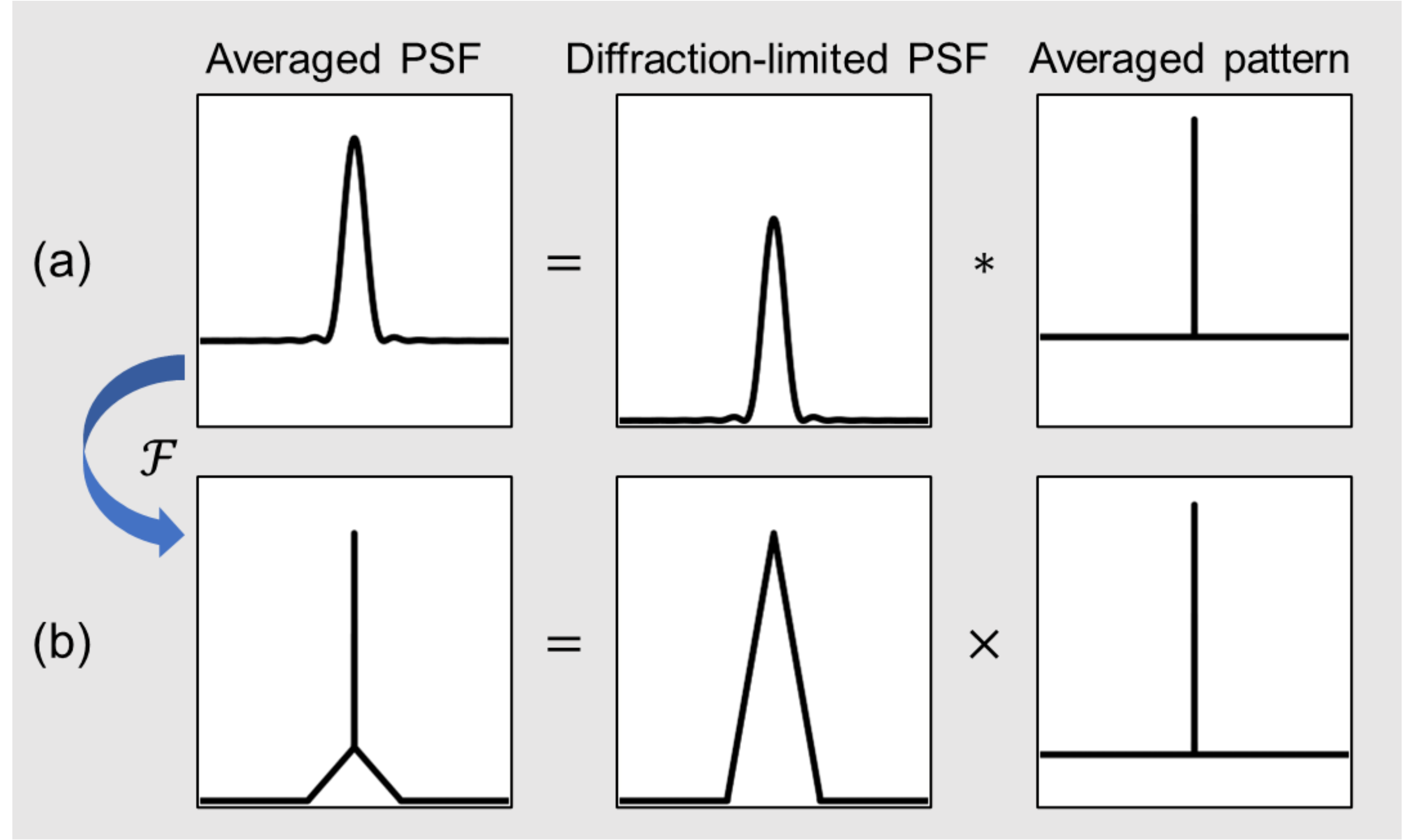}
	\caption{Principle of Fourier phase retrieval. (a) The averaged PSF $ <h_{n,corr}(\boldsymbol{x})> $ is the convolution of the ideal diffraction-limited PSF with the averaged pattern $ <R_{n,corr}(\boldsymbol{x})> $, which is a sharply peaked function on top of a constant offset. (b) In the Fourier domain, the averaged OTF is the multiplication of the triangular function and the sharply peaked with constant offset.}
	\label{fig:fig8}
\end{figure}

While the methods to recover Fourier amplitude are well established, recovery of the Fourier phase has been a more difficult problem. To retrieve the phase, one can apply the Fienup-type iterative phase retrieval method\cite{fienup_reconstruction_1978,fienup_phase_1982} or the speckle masking method\cite{lohmann_speckle_1983,bartelt_phase_1984}, but they either suffer from non-convex optimization challenges or impose excessive computational load due to high-dimensional data processing. Only recently has a simple and effective method called coherent averaging\cite{hwang_imaging_2023} emerged to address the phase retrieval problem. Its principle is illustrated in Fig. \ref{fig:fig8} and simply described here. To start with, the $n$th distorted kernel $h_{n}(\boldsymbol{x})$ that is randomly generated by turbulence can be represented as:
\begin{equation}
	h_{n}(\boldsymbol{x}) = h_{ideal}(\boldsymbol{x})\otimes R_{n}(\boldsymbol{x}).
	\label{eq:eq5}
\end{equation}
where $h_{ideal}(\boldsymbol{x})$ is the ideal diffraction-limited PSF and $R_{n}(\boldsymbol{x})$ is the $n$th instantaneous random position and weights caused by random distortion. The coherent averaging of $h_{n}(\boldsymbol{x})$ can be expressed as:
\begin{equation}
	\begin{split}
		<h_{n,corr}(\boldsymbol{x})> &=h_{ideal}(\boldsymbol{x})*<R_{n,corr}(\boldsymbol{x})>\\
		&=h_{ideal}(\boldsymbol{x})+\mathrm{constant\ background},
	\end{split}
	\label{eq:eq6}
\end{equation}
where $ h_{n,corr}(\boldsymbol{x}) $ is the shift corrected version of $h_{n}(\boldsymbol{x})$ with its maximum value at the origin. The background haze causes ensemble-averaged shift-and-add images to appear diffuse and blurry, limiting adoption of the shift-and-add method despite its simplicity. However, this issue stems from past practices treating such images as final outputs. Instead, since both $h_{ideal}(\boldsymbol{x})$ and the constant background are real-valued and symmetric, the Fourier transform of $ <h_{n,corr}(\boldsymbol{x})> $ is purely real, resulting in a zero-phase Fourier spectrum. Moreover, as shown in Fig. \ref{fig:fig8}(b), the Fourier amplitude of $ <h_{n,corr}(\boldsymbol{x})> $ has nonzero values up to the diffraction limit. Therefore, combining with Equation (2), it can be seen that the Fourier phase of an object can be directly retrieved through simply coherent averaging randomly distorted images:
\begin{equation}
	\phi(\boldsymbol{u}) = \angle\mathcal{F}\{I_{avg}(\boldsymbol{x})\},
	\label{eq:eq7}
\end{equation}
where $\phi(\boldsymbol{u}) $ is the retrieved phase and $I_{avg}(\boldsymbol{x})$ is the averaging of the shift-corrected $\{I_{n}(x)\}$. Combining Equation (\ref{eq:eq6}) and Equaiton (\ref{eq:eq7}) leads to the turbulence-mitigated image:
\begin{equation}
	I_{post}(\boldsymbol{x}) = \mathcal{F}^{-1}\{<|\tilde{I}(\boldsymbol{u})|>\cdot \mathrm{exp}(i\phi(\boldsymbol{u}))\}.
	\label{eq:eq8}
\end{equation}

Notably, the critical step enabling turbulence mitigation for FP is the recovery of the Fourier phase $\phi(\boldsymbol{u})$ via Equation (\ref{eq:eq7}). This phase recovery, combined with the Fourier amplitude obtained from Labeyrie's method, allows the reconstruction of the diffraction-limited image $ I_{post}(\boldsymbol{x}) $ for each sub-aperture. These turbulence-corrected, diffraction-limited sub-aperture images are then fed into the standard FP reconstruction engine to synthesize the final high-resolution image.

\subsection{Reconstruction framework}

\begin{figure}[ht]
	\centering
	\includegraphics[width=10cm]{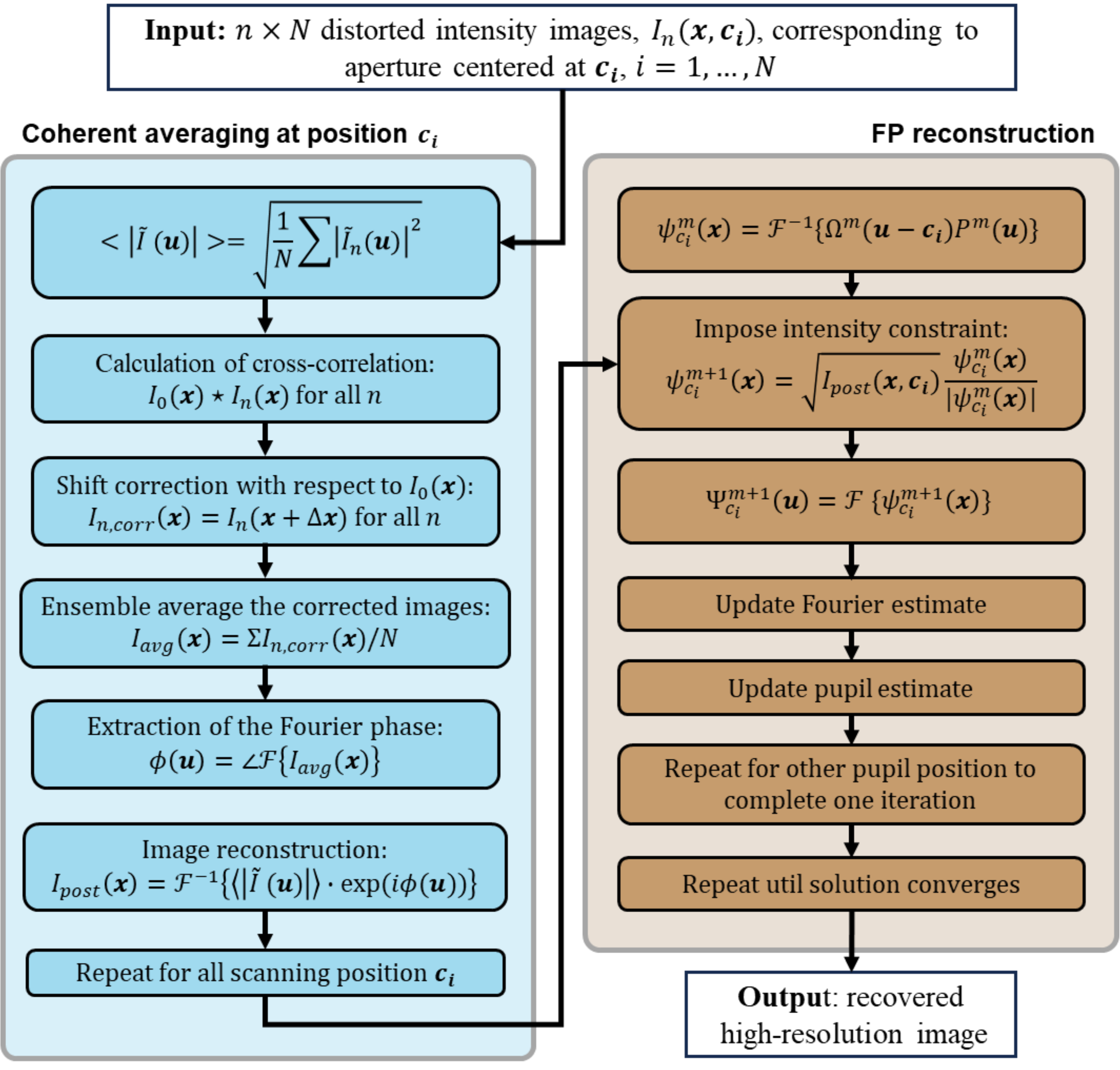}
	\caption{Reconstruction flowchart of TMFP. Coherent averaging recovers the turbulence-reduced images which are then used for FP reconstruction for computational phase retrieval until solution converges, yielding the high-resolution recovery.}
	\label{fig:fig2}
\end{figure}

Figure \ref{fig:fig2} outlines the reconstruciton framework of TMFP with a flow chart, which is detailed as follows.

For turbulence mitigation of each sub-aperture image, the steps are as follow: (1) The orginal sharp image's Fourier amplitude is recovered via Labeyrie's method, which ensemble-averages the power spectra $<|\tilde{I}_n(\boldsymbol{u})|^2>$ of multiple turbulence-distorted images. (2) A reference image $I_{0}(\boldsymbol{x})$ is selected, and all subsequent images $I_n(\boldsymbol{x})$ are spatially aligned to $I_{0}(\boldsymbol{x})$ through cross-correlation-based registration. (3) The shift-corrected images $I_{n,corr}(\boldsymbol{x})$ are ensemble averaged to obtain $I_{avg}(\boldsymbol{x})$, from which the Fourier phase $\phi(\boldsymbol{u})$ is extracted via $\phi(\boldsymbol{u})=\angle\{\mathcal{F}\{I_{avg}(\boldsymbol{x})\}\}$. (4) The extracted Fourier phase $\phi(\boldsymbol{u})$ is combined with the Fourier amplitude $<|\tilde{I}(\boldsymbol{u})|>$, and an inverse Fourier transform yields the turbulence-mitigated result $I_{post}(\boldsymbol{x})$. One can see that the workflow only consists of two Fourier transform and additions of acquired images, and is free of pre-assumed constraints and initial guesses, thus offering a simple and fast Fourier phase retrieval method for computational imaging through random media. Steps (1)-(4) are repeated for all aperture positions, the results are then fed into the FP reconstruction algorithm.

The core of FP is phase retrieval, which allows the recovery of missing phase through the constraint imposed by the overlapping sub-aperture spectra. The FP reconstruction works in an iterative manner, in which constraints are alternately imposed in the spatial and frequency domain. In the $m$-th iteration, the estimate of complex object spectrum and pupi function are denoted as $\Omega^{m}(\boldsymbol{u})$ and $ P^{m}(\boldsymbol{u}) $, respectively. For each camera position $\boldsymbol{c}$, the intensity constraint is enforced in the spatial domain:
\begin{equation}
	\psi_{c}^{m}(\boldsymbol{x})=\mathcal{F}^{-1}\{\Omega_{c}^{m}(\boldsymbol{u}-\boldsymbol{u_c})P^{m}(\boldsymbol{u}) \}
\end{equation}
\begin{equation}
	\psi_{c}^{m+1}(\boldsymbol{x})=\sqrt{I_{post}(\boldsymbol{x})}\frac{\psi_{c}^{m}}{|\psi_{c}^{m}|}
\end{equation}
\begin{equation}
	\Psi_{c}^{m+1}(\boldsymbol{x})=\mathcal{F}\{\psi_{c}^{m+1}(\boldsymbol{x})\}
\end{equation}

In the frequency domain, the estimate of $\Omega(\boldsymbol{u})$ and $P(\boldsymbol{u})$ are updated according to the following formulation:
\begin{equation}
	\begin{split}
		\Omega_{c}^{m+1}(\boldsymbol{x})&=\Omega_{c}^{m}(\boldsymbol{x})+\alpha\frac{[P^{m}(\boldsymbol{u}+\boldsymbol{u_c})]^{*}}{|P^{m}(\boldsymbol{u}+\boldsymbol{u_c})|_{max}^{2}}\\&
		\times [\Psi_{c}^{m+1}(\boldsymbol{x})-\Psi_{c}^{m}(\boldsymbol{x})],
	\end{split}
\end{equation}

\begin{equation}
	\begin{split}
		P^{m+1}(\boldsymbol{x})&=P^{m}(\boldsymbol{x})+\beta\frac{[\Omega^{m}(\boldsymbol{u}-\boldsymbol{u_c})]^{*}}{|\Omega^{m}(\boldsymbol{u}-\boldsymbol{u_c})|_{max}^{2}}\\&
		\times [\Psi_{c}^{m+1}(\boldsymbol{x})-\Psi_{c}^{m}(\boldsymbol{x})],
	\end{split}
\end{equation}
where $\alpha$ and $\beta$ are constants that adjust the step size of update, their values are set to 1 in our work. The process is executed for other pupil positions, wherein all the turbulence-mitigated results are utilized to update both the pupil and object spectra. Upon completion of these updates, a single iteration of the algorithm is deemed concluded. The iterative procedure is continued for multiple iterations until the algorithm achieves a predetermined level of convergence. Subsequently, the object spectrum is inverse Fourier transformed to the spatial domain, thereby obtaining a high-resolution turbulence-mitigated reconstruction. However, due to inherent optical roughness and random phase characteristics exhibited by real-world objects, the FP reconstruction exhibits intensity fluctuations in the speckle pattern, commonly referred to as ‘speckle noise’. To address this artifact, a post-processing denoising step employing a state-of-the-art algorithm recently developed by Tian \emph{et al.}\cite{tian_optical_2023} is implemented on the instant output of TMFP results.

\section{Simulations}

In this section, we use simulation to benchmark our method and characterize its performance. We first simulated imaging scenarios with different degrees of turbulence and compared our proposed method with traditional FP methods. Then, we analyzed the sensitivity of our method to input distorted image count to provide guidance for the application of this method.

\subsection{Feasibility and performance}
\begin{figure}[ht!]
	\centering
	\includegraphics[width=12cm]{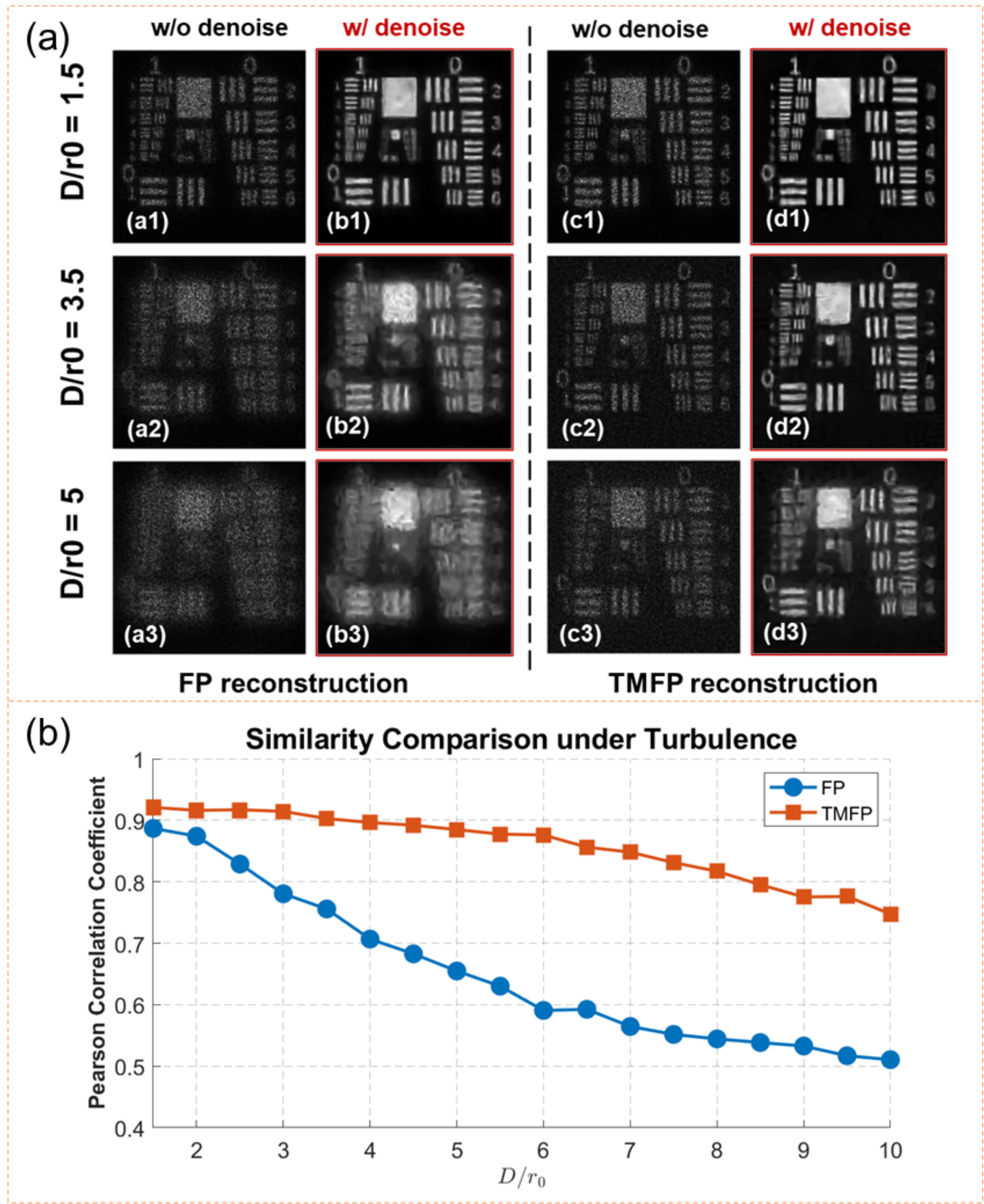}
	\caption{(a) Comparative analysis of FP and TMFP under varying turbulence intensities ($D/r_0$). Rows: different turbulence intensities (up to down: $D/r_{0}=1.5, 3.5, 5$, corresponding to mild, moderate, and severe turbulence). Columns: (a1-a3) FP without denoising: Progressive resolution degradation with increasing $D/r_0$, exhibiting obvious spatial-frequency aliasing. (b1-b3) FP with denoising: Speckle suppression improves contrast but fails to recover high-frequency details under medium and strong turbulence ($D/r_{0} = 3.5$ and $D/r_{0} = 5$). (c1-c3) TMFP without denoising: Reconstruction from turbulence-mitigated results preserves structural integrity even at $D/r_{0} =5$, though residual speckle noise persists. (d1-d3) TMFP with denoising: Combined turbulence mitigation and speckle denoising preserve structural integrity and improve image contrast across all turbulence levels. (b) Pearson correlation coefficient (PCC) comparison between Fourier ptychography (FP) and Coherent Averaging Fourier Ptychography (TMFP) under varying turbulence intensities ($D/r_{0}$)}
	\label{fig:fig3}
\end{figure}

To test the feasibility and performance of the proposed TMFP method with respect to turbulence conditions, we first perform simulations based on phase screen method with Zernike polynomials (see Appendix A), in which random PSFs are generated by applying random phase screens based on Kolmogorov's turbulence model\cite{harding_fast_1999,noll_zernike_1976}. The degree of turbulence is quantified using the parameter $D/r_0$, where $D$ is the diameter of system aperture and $r_0$ is known as the Fried parameter that describes the turbulence characteristic length\cite{fried_optical_1966}. Resolving power is limited by the imaging system itself only when $D<r_0$, in other cases it is limited by the turbulence. Therefore, the larger the value of $D/r_0$, the more severe the turbulence and the lower the imaging resolution.

In our simulation, a $256\times 256$-pixel USAF resolution target with a pixel pitch of $0.6 \mu m$ serves as the object’s amplitude, while its phase is randomized between $-\pi$ to $\pi$ to emulate an optically rough surface. The optical system operates at a wavelength of 632 nm with an F-number of 8, yielding a numerical aperture of 0.0625. A synthetic aperture is formed by simulating a 21×21 grid of camera positions, ensuring enough Fourier domain overlap between adjacent measurements to guarantee algorithmic convergence. At each position, 10 turbulence-distorted short-exposure images are generated, modeling dynamic wavefront aberrations via Kolmogorov phase screens with $D/r_0$ values ranging from 1.5 to 5.0. This results in a total dataset of 4,410 raw images, which are subsequently downsampled by a factor of 4. 

For TMFP reconstruction of the dataset, an initial step involves performing coherent averaging on the 10 distorted images at each measurement position to obtain a turbulence-robust intensity estimation. In contrast, traditional FP reconstruction employs a stochastic selection mechanism where a single distorted image is randomly chosen from each position to form the basis of reconstruction. Notably, both algorithms utilize identical FP processing parameters, including 10 iterative reconstruction steps. The resultant image quality comparisons are under varying turbulence intensities are presented in Figure \ref{fig:fig3}a. For mild turbulence ($D/r_{0} = 1.5$, Figure \ref{fig:fig3}(a1)-(d1)), both methods achieve near-diffraction-limited resolution, and it can be seen that the denoising step indeed increase the image quality. As turbulence intensifies $(D/r_{0} = 3.5$, Figure \ref{fig:fig3}(a2)-(d2)), FP reconstructions degrade significantly, whereas TMFP maintains fidelity. Under extreme turbulence ($D/r_{0} =5$, Figure \ref{fig:fig3}(a3)-(d3)), FP fails to resolve critical features, while TMFP retains structural integrity with only moderate high-frequency attenuation.

For quantitative analysis, we use Pearson correlation coefficient (PCC) as the similarity score to evaluate the reconstruction performance, which is given in Appendix B. As shown in Figure \ref{fig:fig3}b, the similarity scores of both FP and TMFP decrease with increasing turbulence intensity ($D/r_0$), but TMFP consistently outperforms FP across all test conditions. Under mild turbulence ($D/r_{0}=1.5$), TMFP achieves a PCC of $0.9209$, slightly higher than FP’s $0.8867$. However, as turbulence intensifies, the gap widens significantly: at $D/r_{0} = 5$, TMFP maintains a PCC of $0.8848$, while FP drops to $0.6548$. Under severe turbulence ($D/r_{0}=10$), TMFP’s PCC ($0.7475$) remains $46\%$ higher than FP’s ($0.5105$), demonstrating its superior robustness to turbulence-induced distortions. These quantitative metrics align with the visual results in Figure \ref{fig:fig3}a, where TMFP preserves structural details even under severe turbulence, while FP suffers from spatial-frequency aliasing and pseudo-resolution artifacts. Collectively, these findings validate TMFP’s capability to maintain high-fidelity imaging in dynamic scattering environments.

\subsection{Sensitivity to distorted image count}

The robustness of TMFP against varying numbers of turbulence-distorted input images per scanning position is critical for practical deployment. To quantify this relationship, we evaluated reconstruction fidelity under different turbulent conditions using input image counts ranging from $N = 5$ to $N = 100$ per position. 

\begin{figure}[tb]
	\centering
	\includegraphics[width=12cm]{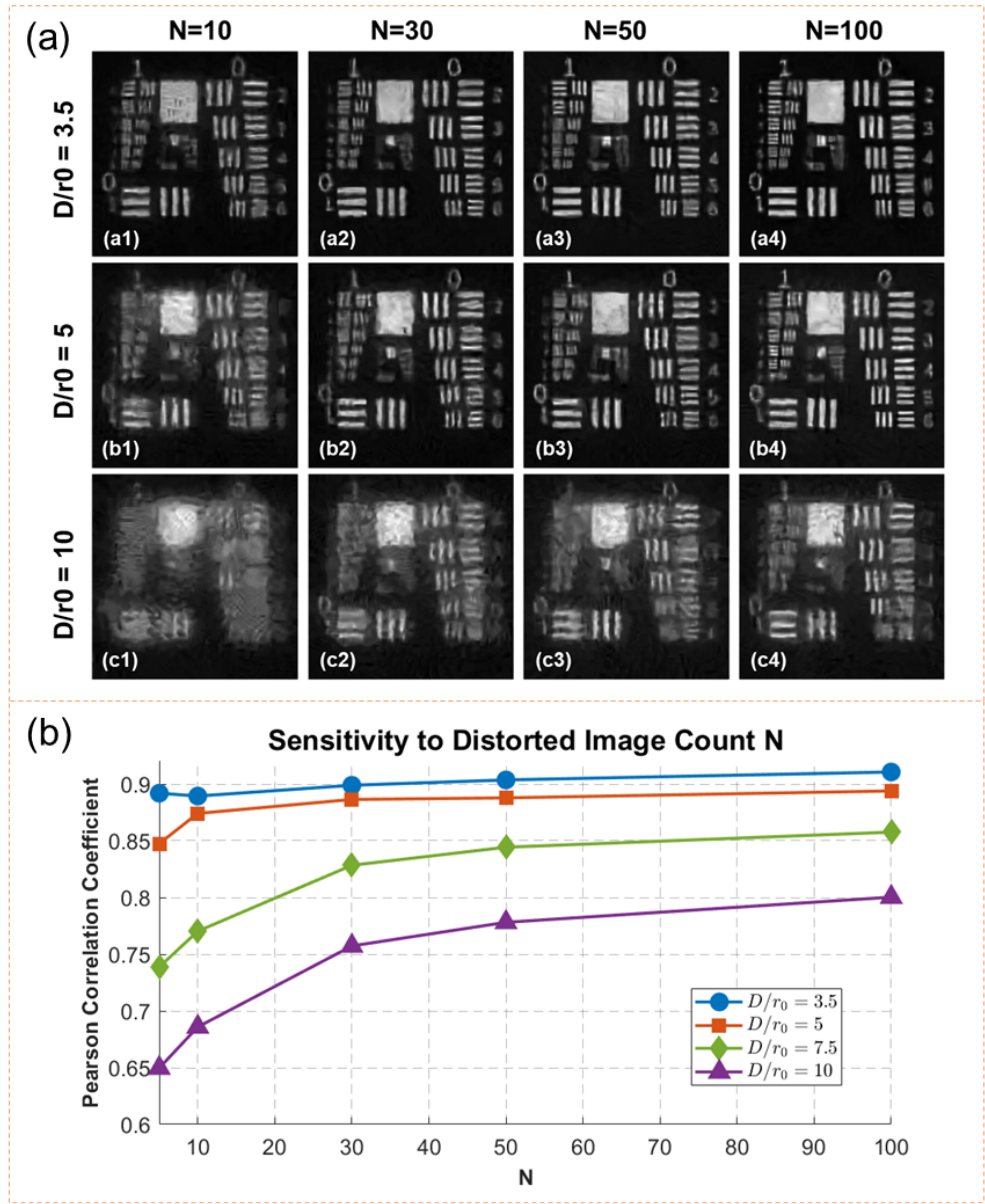}
	\caption{(a) Representative visual results. Rows: Turbulence regimes (up to down:  $D/r_{0}=3.5$, $D/r_{0}=5$, and $D/r_{0}=10$). Columns: Input image counts per scanning position (left to right: $10, 30, 50, 100$), demonstrating progressive improvement in turbulence compensation. (b) Sensitivity analysis of TMFP reconstruction fidelity quantified by PCC to the number of turbulence-distorted input images ($N$) across turbulence intensities.}
	\label{fig:fig4}
\end{figure}

Figure \ref{fig:fig4}(b) quantifies the reconstruction performance of TMFP using the PCC as a function of the turbulence-distorted input image count $N$, with representative visual results shown in Figure \ref{fig:fig4}a. One can see that TMFP’s resolution improves asymptotically with $N$. Turbulence severity governs the required $N$ for artifact suppression and detail recovery: under moderate turbulence for artifact suppression and detail recovery: under moderate turbulence ($D/r_{0} = 3.5$), near-diffraction-limited resolution is achieved with as few as $N=5$, while increasing $N$ to $100$ yields negligible gains. For stronger turbulence ($D/r_{0} = 5$), $N\geq30$ is necessary to effectively mitigate spatial-frequency aliasing and restore high-frequency features. Under even stronger turbulence ($D/r_{0} = 7.5$ and $10$), reconstruction fidelity degrades markedly, necessitating $N\geq50$ and $N\geq100$, respectively, to achieve acceptable results.

While higher $N$ enhances turbulence suppression, it linearly increases data acquisition time and computational load. For example, processing $N = 50 $ images per position extends reconstruction time by $5.4\times$ compared to $N = 10$, with diminishing returns in resolution improvement. Therefore, it is essential to find a sweet spot that balances fidelity and efficiency. To sum up, in scenarios with severe turbulence, increasing $N$ to $50$ and larger is necessary to mitigate artifacts caused by turbulence distortion. For moderate or weakly turbulent conditions, $N =10$ suffices to achieve near-optimal performance. These results provide guideline for adaptive resource allocation in real-world systems, such as aerial surveillance, where data throughput and latency are critical constraints.

\section{Experiments}
\begin{figure}[ht]
	\centering
	\includegraphics[width=10cm]{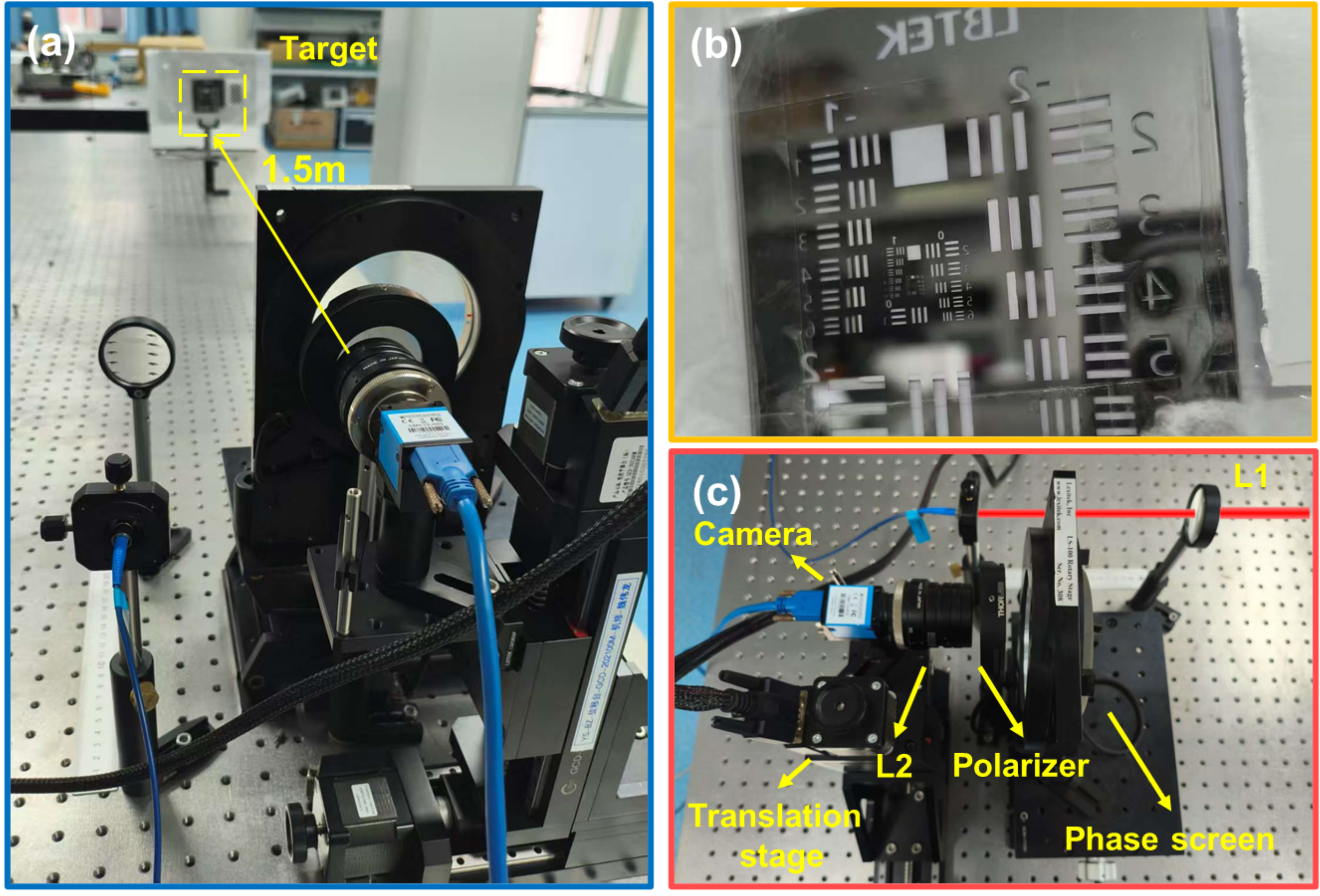}
	\caption{TMFP for macroscale photographic imaing. (a) The experimental setup, where the target is placed 1.5 m away from the imaging system. (b) The USAF resolution target. (c) The illumination and detection system.}
	\label{fig:fig5}
\end{figure}

We conduct proof-of-concept experiment to demonstrate the performance of the proposed method. As illustrated in Figure \ref{fig:fig5}, the experimental setup employs a 532 $nm$ laser source coupled to a single-mode fiber and collimated through a plano-convex lens (L1, focal length: 200 $mm$) to produce the illumination beam. The target is positioned 1.5 $m$ away from the detection system. In the detection arm, a rotatable phase screen is placed in front of the camera to generate random turbulence distortion. A linear polarizer filters noninterference background light. Light from the target is collected using a photographic lens (L2, focal length: 50 $mm$, $f/8$ aperture) and detected by a monochrome camera (Imagingsource, 2.4 $\mu m$ pixel pitch). Two motorized stages are used to raster-scan the camera to different x-y positions with a step size of 1.5 $mm$, resulting in $85\%$ Fourier domain overlap between adjacent positions. At each scanning position, the phase screen is rotated to multiple angles to capture degraded images under different turbulence disturbances. Notably, the phase screen is placed solely in the detection arm, which means that we only consider the distortion on the reflection wavefront. This is reasonable because the effect of turbulence distortion on illumination wavefront is limited as it can be overwhelmed by the random phase of rough objects.

In the first experiment, we use a USAF resolution test chart as the target, as shown in Figure \ref{fig:fig5}b. The camera is raster-scanned across a $11\times 11$ grid, synthesizing an effective aperture of 21.35 $mm$, which is about $3.4\times$ larger than the single aperture (6.25 $mm$). In each scanning position, the phase screen is rotated manually and 10 distorted images is recorded, resulting in a dataset containing a total of 1210 turbulence-degraded images ($11\times 11$ positions $\times 10$ turbulence realizations, visualized in Figure \ref{fig:fig1}b). For benchmarking purposes, a turbulence-free dataset is acquired by removing the phase screen, while ensuring identical scanning parameters and illumination conditions.

\begin{figure}[ht]
	\centering
	\includegraphics[width=10cm]{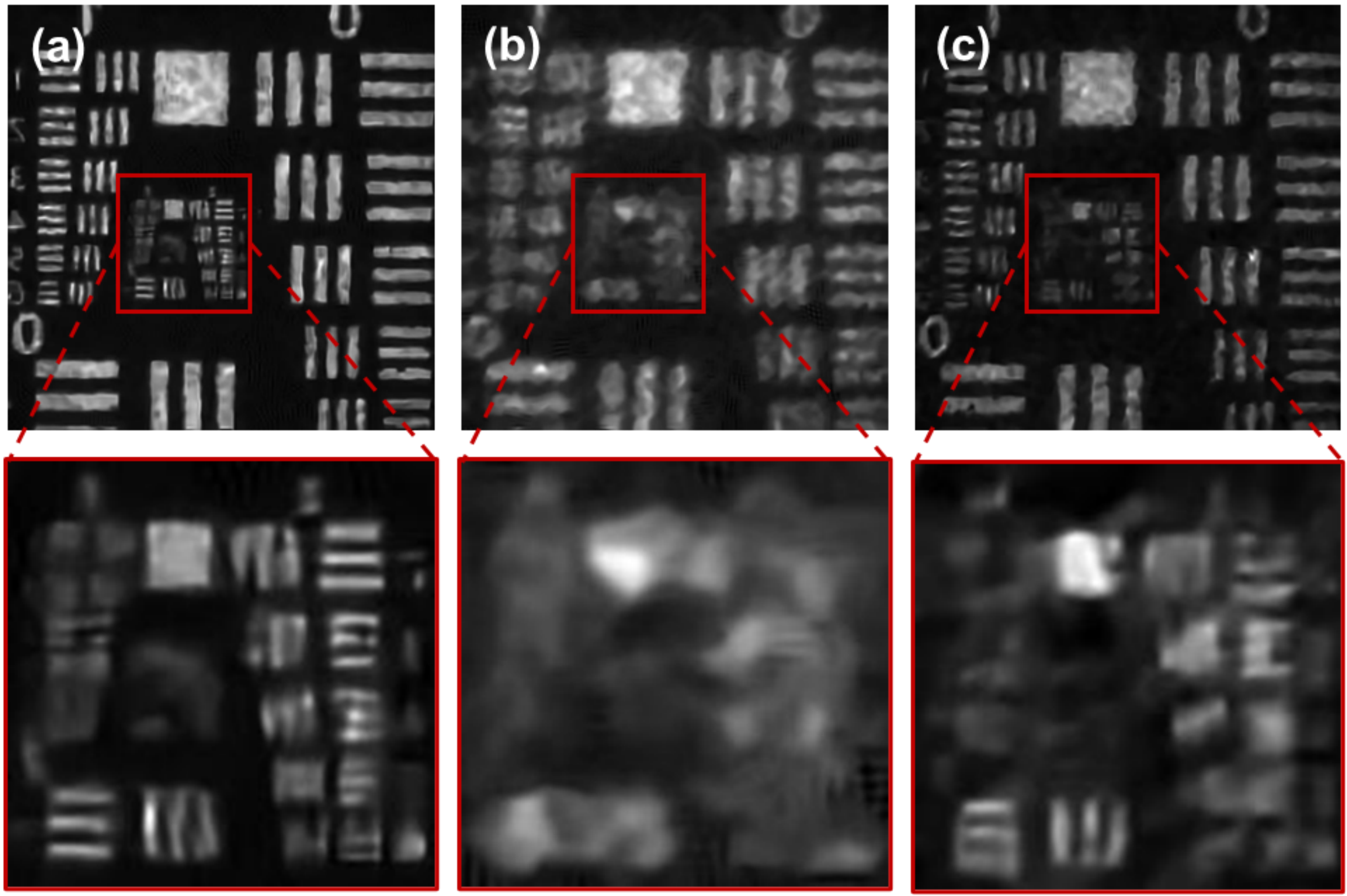}
	\caption{Experimental results of a USAF resolution target. (a) FP reconstruction under ideal condition. (b) FP reconstruction under turbulent condtion. (c) TMFP reconstruction under turbulent condition.}
	\label{fig:fig6}
\end{figure}

Figure \ref{fig:fig6}a shows the reconstruction result of the turbulence-free dataset, from which we can see the maximum resolvable line pair is Group 2 Element 4, corresponding to an angular resolution of $5.89\times 10^{-5}$ rad. For coherent imaging system operating under speckle conditions, the theoretical Rayleigh resolution limit is given by $R=1.6\lambda/D$. Substituting the wavelength $\lambda = 639 \,nm$ and the synthetic aperture size $D = 21.35\,mm$, the calculated theoretical angular resolution is $4.79\times10^{-5}$ rad, which aligns with the experimentally measured value. Figure \ref{fig:fig6}b presents the FP reconstruction results of the turbulence-degraded dataset, where only a single turbulence-corrupted image per scanning position is utilized for recovery. The turbulence-induced distortions severely degrade the imaging performance, reducing the maximum resolvable feature from Group 2 Element 4 (turbulence-free case) to Group 1 Element 1 on the USAF target—a threefold reduction in spatial resolution. The significant resolution loss underscores the limitations of conventional single-frame FP reconstruction under turbulence and highlights the necessity of multi-frame turbulence compensation mechanisms. Figure \ref{fig:fig6}c demonstrates the TMFP reconstruction results, showcasing a significant enhancement in both resolution and image quality compared to Figure \ref{fig:fig6}b. TMFP restores the maximum resolvable feature to Group 2 Element 2 on the USAF target—a twofold improvement over the turbulence-corrupted result (Group 1 Element 1 in Figure \ref{fig:fig6}b). Quantitatively, this corresponds to an angular resolution enhancement from $1.75\times10^{-4}$ rad to $8.75\times10^{-5}$ rad, approaching $80\%$ of the turbulence-free performance.

\begin{figure}[ht]
	\centering
	\includegraphics[width=10cm]{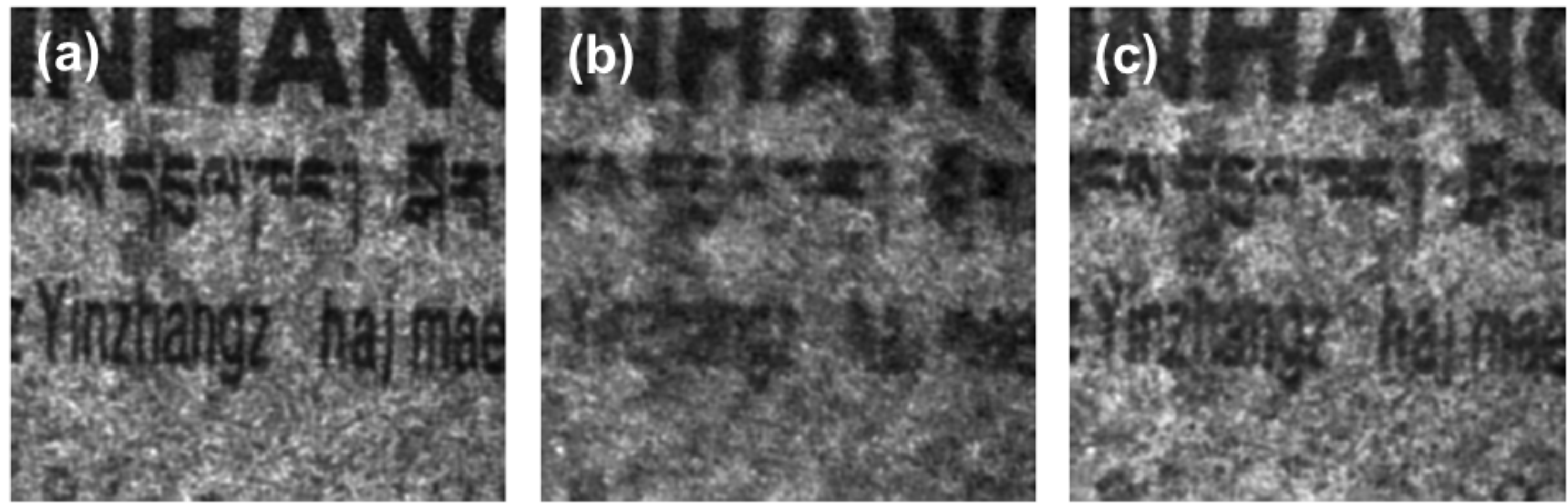}
	\caption{Experimental results of the text part on a five-yuan banknote. (a) FP reconstruction under ideal condition. (b) FP reconstruction under turbulent condtion. (c) TMFP reconstruction under turbulent condition.}
	\label{fig:fig7}
\end{figure}

In the second experiment, we use a five-yuan RMB as the target. Unlike the binary intensity profile of the USAF resolution test chart, the banknote exhibits continuous grayscale variations and stronger phase scattering effects, leading to significantly reduced image contrast and more pronounced speckle noise. The reconstruction results demonstrate TMFP's effectiveness: (a) the turbulence-free reconstruction in Figure \ref{fig:fig7}a clearly resolves all text elements with high fidelity; (b) the turbulence-degraded single-frame FP reconstruction in Figure \ref{fig:fig7}b fails to distinguish the second and third text lines due to severe phase distortions; while (c) the TMFP reconstruction in Figure \ref{fig:fig7}c successfully restores text legibility through multi-frame turbulence compensation, achieving performance comparable to the turbulence-free case.

\section{Conclusion}

In summary, we have demonstrated a robust turbulence-mitigated Fourier ptychography for active high-resolution imaging through optical turbulence. By considering the turbulence distortion as an image degradation process, the state-of-art computational methods can be applied to mitigate the distortion of optical turbulence, our TMFP method achieves high-fidelity results under non-idealized conditions without altering the optical setup. The simulative results show that our method effectively operates under strong turbulence regimes ($D/r_0 = 5$), while for mild to moderate turbulence, a balance is achieved between algorithmic performance and computation load effectively operates under strong turbulence regimes $N = 10$ distorted input images for coherent averaging. Our $1.5m$ proof-of-principal experiment further demonstrate the superiority of this method, achieving $80\%$ of the resolution fidelity achievable in turbulence-free scenarios, and achieving twice the resolution improvement and significant image quality improvement compared to traditional FP methods. 

Overall, we believe this approach establishes a new paradigm for high-fidelity imaging in adverse environments, achieving an organic integration of active illumination super-resolution imaging and computational turbulence correction. Future work will focus on extending this framework to practical applications, optimizing computational workflows for large-scale datasets, and exploring synergies with adaptive optics for even greater resilience in extreme turbulence scenarios. This approach holds promise for applications in free-space optical communication, astronomical imaging, and remote sensing, where turbulence compensation is critical.

\section*{Appendix A: Phase screen method based on Zernike polynomial}
\setcounter{equation}{0}
\renewcommand{\theequation}{A\arabic{equation}}

The phase screen method based on Zernike polynomial is a widely-used numerical approach to simulate the effect of turbulence, in which random phases are produced over the pupil and the atmospheric correlation is introduced with proper filtering by the Kolmogoroff spectrum. The phase screen is represented by a linear combination of a series of Zernike polynomial\cite{noll_zernike_1976}:
\begin{equation}
	\phi = \sum_{j=1}^{\infty}a_{j}Z_{j},
	\label{eq:eqc1}
\end{equation}
where the phase screen is represented by $\phi$, $Z_{j}$ is the $j$-th order Zernike polynomial, $a_{j}$ is the Zernike coefficient which is obtained by calculating its covariance matrix due to its atmospheric correlation. For detailed explanation, the reader can refer to Noll’s work\cite{noll_zernike_1976}, here we directly give the expression of $(i,j)$ element of the covariance matrix when $i-j=2k$ (when $i-j=2k+1$, $ <a_ia_j>=0 $):
\begin{equation}
	\begin{split}
		&<a_ia_j>=0.15(D/r_0)^{\frac{5}{3}}\cdot \\&
		\frac{(-1)^{\frac{n_i+n_j-2m_i}{2}}[(n_i+1)(n_j+1)]^{1/2}\Gamma(\frac{14}{3})\Gamma[(n_i+n_j-\frac{5}{3})/2]\delta_{z}}{\Gamma(\frac{n_i-n_j+17/3}{2})\Gamma(\frac{n_j-n_i+17/3}{2})\Gamma(\frac{n_i+n_j+23/3}{2})},
	\end{split}
	\label{eq:eqc2}
\end{equation}
where $n_i$, $n_j$, $m_i$, $m_j$ are the Noll coefficients corresponding to the $i$-th and $j$-th orders of Zernike polynomials, $D$ is the aperture size, $r_0$ is the coherence length, and $\delta_z$ is the logical Kronecker symbol. Then, singular value decomposition is applied to the covariance matrix, and the coefficient $a_j$ can be calculated by introducing Gaussian random variables $b$ according to its eigenvector $U$:
\begin{equation}
	a_{j} = \sum_{k=1}^{\infty}U_{jk}b_{j}.
	\label{eq:eqc3}
\end{equation}

The phase screen is then obtained by adding together Zernike polynomials of each mode according to Equation (\ref{eq:eqc1}). Finally, the resultant pupil function $P(\boldsymbol{u})$ is given by:
\begin{equation}
	P(\boldsymbol{u}) = circ(k_0\cdot \frac{1}{2f/\#})\mathrm{exp}(-j\phi),
	\label{eq:eqc4}
\end{equation}
where $circ(\cdot)$ denotes a circular mask with a radious of $k_0\cdot \frac{1}{2f/\#}$, with $k_0$ representing the wave number. Notably, various turbulence conditoins can be simulated by adjusting the order of Zernike polynomials and the value of $D/r_0$ in Equation (\ref{eq:eqc2}). In this work, we use the first 15 orders of Zernike polynomials to generate the phase screen. The whole process is summarized as follows:

1. In each camera position, randomly generate $N$ phase screens $\phi_{n}$, $n = 1,2,...,N$, according to the above process, where $N$ is the number of short-exposure images.

2. Calculate the corresponding turbulence-distorted pupil function $P_n(\boldsymbol{u})$ according to Equation (\ref{eq:eqc4}).

3. Obtain the $N$ turbulence-distorted images according to Equation (\ref{eq:eq1}).

4. Repeat the above process for all camera position.

\section*{Appendix B: Pearson Correlation Coefficient}
\setcounter{equation}{0}
\renewcommand{\theequation}{B\arabic{equation}}
The similarity between TMFP reconstruction and ground truth can be evaluated by the Pearson correlation coefficient (PCC), which is given by:
\begin{equation}
	r=\dfrac{\sum_{m}\sum_{n}(A_{mn}-\overline{A})(B_{mn}-\overline{B})}{\sqrt{(\sum_{m}\sum_{n}(A_{mn}-\overline{A})^2)(\sum_{m}\sum_{n}(B_{mn}-\overline{B})^2)}},
\end{equation}
where $r$ is the value of PCC, $A$ and $B$ are two-dimensional matirx representing the reconstruction result and the ground truth, $A_{mn}$ denotes the elements in matrix $A$, $\overline{A}$ denotes the average value of $A$.

\begin{backmatter}
	\bmsection{Funding} This work was supported by the National Natural Science Foundation of China (Grant No. 62175243); the Youth Innovation Promotion Association, CAS (Grant No. 2020372); and Sichuan Science and Technology Program (Grant No. 2023YFSY0059).
	
	\bmsection{Acknowledgments} Junhao Zhang thanks Dr. Di You for the helpful discussion, as well as Dr. Yanda Gu and Dr. Yingben Song for their assistance in experimental data collection.
	
	\bmsection{Disclosures} The authors declare no conflicts of interest.

	\bmsection{Data Availability} Data underlying the results presented in this paper are not publicly available at this time but may be obtained from the authors upon reasonable request.

\end{backmatter}


\bibliography{citation_list}

\begin{thebibliography}{10}
\newcommand{\enquote}[1]{``#1''}

\bibitem{liu_research_2025}
Q.~Liu, Y.~Di, M.~Zhang, \emph{et~al.}, \enquote{Research {Progress} on
  {Atmospheric} {Turbulence} {Perception} and {Correction} {Based} on
  {Adaptive} {Optics} and {Deep} {Learning},} {\protect\JournalTitle{Advanced
  Photonics Research}} p. 2400204 (2025).

\bibitem{miller_optical_2007}
N.~J. Miller, M.~P. Dierking, and B.~D. Duncan, \enquote{Optical sparse
  aperture imaging,} {\protect\JournalTitle{Appl. Opt.}} \textbf{46},
  5933--5943 (2007). Publisher: Optica Publishing Group.

\bibitem{moreira_tutorial_2013}
A.~Moreira, P.~Prats-Iraola, M.~Younis, \emph{et~al.}, \enquote{A tutorial on
  synthetic aperture radar,} {\protect\JournalTitle{IEEE Geoscience and Remote
  Sensing Magazine}} \textbf{1}, 6--43 (2013).

\bibitem{ryle_synthesis_1960}
M.~Ryle, A.~Hewish, and S.~R., \enquote{The synthesis of large radio telescopes
  by the use of radio interferometers,} {\protect\JournalTitle{Antennas and
  Propagation, IRE Transactions on}} \textbf{7}, 120 -- 124 (1960).

\bibitem{xie_15-m_2023}
Z.~Xie, K.~Yang, Y.~Liu, \emph{et~al.}, \enquote{1.5-m flat imaging system
  aligned and phased in real time,} {\protect\JournalTitle{Photonics Research}}
  \textbf{11}, 1339 (2023).

\bibitem{yang_model-driven_2024}
K.~Yang, X.~Ma, W.~Wei, \emph{et~al.}, \enquote{Model-driven extended scene
  piston sensing for synthetic aperture telescopes,}
  {\protect\JournalTitle{Opt. Express}} \textbf{32}, 42071--42090 (2024).
  Publisher: Optica Publishing Group.

\bibitem{zheng_wide-field_2013}
G.~Zheng, R.~Horstmeyer, and C.~Yang, \enquote{Wide-field, high-resolution
  {Fourier} ptychographic microscopy,} {\protect\JournalTitle{Nature
  Photonics}} \textbf{7}, 739--745 (2013).

\bibitem{zheng_concept_2021}
G.~Zheng, C.~Shen, S.~Jiang, \emph{et~al.}, \enquote{Concept, implementations
  and applications of {Fourier} ptychography,} {\protect\JournalTitle{Nature
  Reviews Physics}} \textbf{3}, 207--223 (2021).

\bibitem{ou_quantitative_2013}
X.~Ou, R.~Horstmeyer, C.~Yang, and G.~Zheng, \enquote{Quantitative phase
  imaging via {Fourier} ptychographic microscopy,} {\protect\JournalTitle{Opt.
  Lett.}} \textbf{38}, 4845--4848 (2013). Publisher: Optica Publishing Group.

\bibitem{dong_high-resolution_2014}
S.~Dong, P.~Nanda, R.~Shiradkar, \emph{et~al.}, \enquote{High-resolution
  fluorescence imaging via pattern-illuminated {Fourier} ptychography,}
  {\protect\JournalTitle{Optics Express}} \textbf{22}, 20856 (2014).

\bibitem{nguyen_deep_2018}
T.~Nguyen, Y.~Xue, Y.~Li, \emph{et~al.}, \enquote{Deep learning approach for
  {Fourier} ptychography microscopy,} {\protect\JournalTitle{Optics Express}}
  \textbf{26}, 26470 (2018).

\bibitem{song_synthetic_2022}
P.~Song, S.~Jiang, T.~Wang, \emph{et~al.}, \enquote{Synthetic aperture
  ptychography: coded sensor translation for joint spatial-{Fourier} bandwidth
  expansion,} {\protect\JournalTitle{Photonics Research}} \textbf{10},
  1624--1632 (2022). Publisher: Optica Publishing Group.

\bibitem{zhou_fourier_2023}
H.~Zhou, B.~Y. Feng, H.~Guo, \emph{et~al.}, \enquote{Fourier ptychographic
  microscopy image stack reconstruction using implicit neural representations,}
  {\protect\JournalTitle{Optica}} \textbf{10}, 1679 (2023).

\bibitem{wu_lens-free_2024}
X.~Wu, N.~Zhou, Y.~Chen, \emph{et~al.}, \enquote{Lens-free on-chip {3D}
  microscopy based on wavelength-scanning {Fourier} ptychographic diffraction
  tomography,} {\protect\JournalTitle{Light: Science \& Applications}}
  \textbf{13}, 237 (2024).

\bibitem{zhao_deep-ultraviolet_2024}
Q.~Zhao, R.~Wang, S.~Zhang, \emph{et~al.}, \enquote{Deep-ultraviolet {Fourier}
  ptychography ({DUV}-{FP}) for label-free biochemical imaging via
  feature-domain optimization,} {\protect\JournalTitle{APL Photonics}}
  \textbf{9}, 090801 (2024).

\bibitem{zhang_toward_2024}
J.~Zhang, W.~Wei, Z.~Xie, \emph{et~al.}, \enquote{Toward robust
  super-resolution imaging: {A} low-rank approximation approach for
  pattern-illuminated {Fourier} ptychography,} {\protect\JournalTitle{APL
  Photonics}}  (2024).

\bibitem{dong_aperture-scanning_2014}
S.~Dong, R.~Horstmeyer, R.~Shiradkar, \emph{et~al.}, \enquote{Aperture-scanning
  {Fourier} ptychography for {3D} refocusing and super-resolution macroscopic
  imaging,} {\protect\JournalTitle{Optics Express}} \textbf{22}, 13586 (2014).

\bibitem{holloway_toward_2016}
J.~Holloway, M.~S. Asif, M.~K. Sharma, \emph{et~al.}, \enquote{Toward
  {Long}-{Distance} {Subdiffraction} {Imaging} {Using} {Coherent} {Camera}
  {Arrays},} {\protect\JournalTitle{IEEE Transactions on Computational
  Imaging}} \textbf{2}, 251--265 (2016).

\bibitem{holloway_savi_2017}
J.~Holloway, Y.~Wu, M.~K. Sharma, \emph{et~al.}, \enquote{{SAVI}: {Synthetic}
  apertures for long-range, subdiffraction-limited visible imaging using
  {Fourier} ptychography,} {\protect\JournalTitle{Science Advances}}
  \textbf{3}, e1602564 (2017).

\bibitem{li_farfield_2023}
S.~Li, B.~Wang, K.~Liang, \emph{et~al.}, \enquote{Far‐{Field} {Synthetic}
  {Aperture} {Imaging} via {Fourier} {Ptychography} with {Quasi}‐{Plane}
  {Wave} {Illumination},} {\protect\JournalTitle{Advanced Photonics Research}}
  p. 2300180 (2023).

\bibitem{article}
Q.~Zhang, Y.~Lu, Y.~Guo, \emph{et~al.}, \enquote{200 mm optical synthetic
  aperture imaging over 120 meters distance via macroscopic fourier
  ptychography,} {\protect\JournalTitle{Optics Express}} \textbf{32},
  44252--44264 (2024).

\bibitem{tian_optical_2023}
Z.~Tian, M.~Zhao, D.~Yang, \emph{et~al.}, \enquote{Optical remote imaging via
  {Fourier} ptychography,} {\protect\JournalTitle{Photonics Research}}
  \textbf{11}, 2072 (2023).

\bibitem{wang_learning-based_2023}
B.~Wang, S.~Li, Q.~Chen, and C.~Zuo, \enquote{Learning-based single-shot
  long-range synthetic aperture {Fourier} ptychographic imaging with a camera
  array,} {\protect\JournalTitle{Optics Letters}} \textbf{48}, 263 (2023).

\bibitem{li_snapshot_2024}
S.~Li, B.~Wang, H.~Guan, \emph{et~al.}, \enquote{Snapshot macroscopic {Fourier}
  ptychography: far-field synthetic aperture imaging via illumination
  multiplexing and camera array acquisition,} {\protect\JournalTitle{Advanced
  Imaging}} \textbf{1}, 011005 (2024).

\bibitem{xiang_coherent_2021}
M.~Xiang, A.~Pan, Y.~Zhao, \emph{et~al.}, \enquote{Coherent synthetic aperture
  imaging for visible remote sensing via reflective {Fourier} ptychography,}
  {\protect\JournalTitle{Optics Letters}} \textbf{46}, 29 (2021).

\bibitem{xiang_phase_2022}
M.~Xiang, A.~Pan, J.~Liu, \emph{et~al.}, \enquote{Phase {Diversity}-{Based}
  {Fourier} {Ptychography} for {Varying} {Aberration} {Correction},}
  {\protect\JournalTitle{Frontiers in Physics}} \textbf{10} (2022).

\bibitem{noll_zernike_1976}
R.~J. Noll, \enquote{Zernike polynomials and atmospheric turbulence*,}
  {\protect\JournalTitle{Journal of the Optical Society of America}}
  \textbf{66}, 207 (1976).

\bibitem{harding_fast_1999}
C.~M. Harding, R.~A. Johnston, and R.~G. Lane, \enquote{Fast simulation of a
  kolmogorov phase screen,} {\protect\JournalTitle{Applied Optics}}
  \textbf{38}, 2161 (1999).

\bibitem{fried_optical_1966}
D.~L. Fried, \enquote{Optical {Resolution} {Through} a {Randomly}
  {Inhomogeneous} {Medium} for {Very} {Long} and {Very} {Short} {Exposures},}
  {\protect\JournalTitle{Journal of the Optical Society of America}}
  \textbf{56}, 1372 (1966).

\bibitem{paxman_phase-diversity_1994}
R.~G. Paxman, B.~J. Thelen, and J.~H. Seldin, \enquote{Phase-diversity
  correction of turbulence-induced space-variant blur,}
  {\protect\JournalTitle{Optics Letters}} \textbf{19}, 1231 (1994).

\bibitem{labeyrie_attainment_1970}
A.~Labeyrie, \enquote{Attainment of {Diffraction} {Limited} {Resolution} in
  {Large} {Telescopes} by {Fourier} {Analysing} {Speckle} {Patterns} in {Star}
  {Images},} {\protect\JournalTitle{Astronomy and Astrophysics}} \textbf{6}
  (1970).

\bibitem{denker_near_2001}
C.~Denker, G.~Yang, and H.~Wang, \enquote{Near {Real}-{Time} {Image}
  {Reconstruction},} {\protect\JournalTitle{Solar Physics}} \textbf{202},
  63--70 (2001).

\bibitem{fienup_reconstruction_1978}
J.~R. Fienup, \enquote{Reconstruction of an object from the modulus of its
  {Fourier} transform,} {\protect\JournalTitle{Optics Letters}} \textbf{3}, 27
  (1978).

\bibitem{fienup_phase_1982}
J.~R. Fienup, \enquote{Phase retrieval algorithms: a comparison,}
  {\protect\JournalTitle{Applied Optics}} \textbf{21}, 2758 (1982).

\bibitem{lohmann_speckle_1983}
A.~W. Lohmann, G.~Weigelt, and B.~Wirnitzer, \enquote{Speckle masking in
  astronomy: triple correlation theory and applications,}
  {\protect\JournalTitle{Applied Optics}} \textbf{22}, 4028 (1983).

\bibitem{bartelt_phase_1984}
H.~Bartelt, A.~Lohmann, and B.~Wirnitzer, \enquote{Phase and amplitude recovery
  from bispectra,} {\protect\JournalTitle{Applied Optics}} \textbf{23},
  3121--3129 (1984).

\bibitem{hwang_imaging_2023}
B.~Hwang, T.~Woo, C.~Ahn, and J.~Park, \enquote{Imaging through {Random}
  {Media} {Using} {Coherent} {Averaging},} {\protect\JournalTitle{Laser \&
  Photonics Reviews}} \textbf{17}, 2200673 (2023).

\end{thebibliography}

\end{document}